\documentclass[twocolumn]{revtex4}
\usepackage[dvips]{graphicx}
\usepackage{amsfonts}
\usepackage{amsmath}
\usepackage{float, color,array, multirow}

\begin{document}

\title{Little-Parks oscillation and ${\bf d}$-vector texture in spin-triplet superconducting rings with bias current}

\author{Kazushi Aoyama}

\date{\today}

\affiliation{Department of Earth and Space Science, Graduate School of Science, Osaka University, Osaka 560-0043, Japan}

\begin{abstract}
We theoretically investigate the critical bias current $j_c$ for a superconducting (SC) ring in a magnetic field. Based on the Ginzburg-Landau theory, we show that $j_c$ exhibits a Little-Parks (LP) oscillation as a function of the magnetic flux passing through the ring, similarly to the LP oscillation in the SC transition temperature. It is also found that for a spin-triplet SC ring, the ${\bf d}$-vector rotates to yield a larger $j_c$, forming a texture along the circumference of the ring. Experimental implications of our result are discussed. 
\end{abstract}

\maketitle
{\it Introduction.}--
In superconductors, a phase of the macroscopic wave function describing the Cooper pair condensate, i.e., a phase of the gap function, often plays an important role for electromagnetic properties of the system. A typical example of such a phase-related phenomenon is the Little-Parks (LP) oscillation emerging in a superconducting (SC) ring in a magnetic field, where the phase and an associated SC current discontinuously change with increasing field and resultantly, a SC transition temperature $T_c$ exhibits a quantum oscillation as a function of field with its period being characterized by the flux quantum $\Phi_0=\frac{hc}{2|e|}$ \cite{LP_LittleParks_prl_62,LP_ParksLittle_pr_64,LP_Groff_pr_68}. In this work, we consider a situation where a bias electric current is further applied to the ring, and theoretically investigate the LP oscillation in the critical bias current $j_c$ at a temperature below $T_c$. In contrast to the spin-singlet case where only the phase degree of freedom is active, in the spin-triplet case, not only the phase but also the ${\bf d}$-vector degrees of freedom characterizing the three spin states turn out to be essential for $j_c$.

\begin{figure}[t]
\begin{center}
\includegraphics[width=\columnwidth]{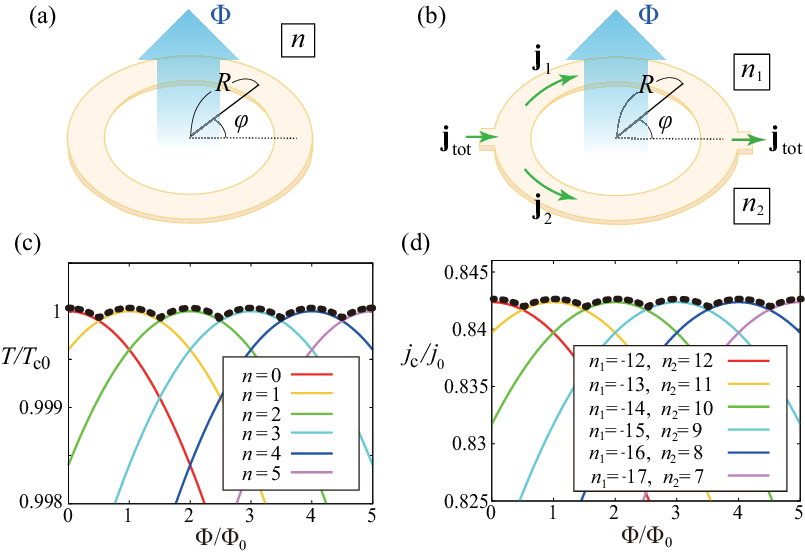}
\caption{System setup and the Little-Parks (LP) oscillation. (a) and (b) SC rings of mean radius $R$ involving the magnetic flux $\Phi$ in the cases without (a) and with (b) a bias current ${\bf j}_{\rm tot}={\bf j}_1+{\bf j}_2$, where $\varphi$ denotes an azimuthal angle. In (a), $n$ is a phase winding number for the whole system, whereas in (b), $n_1$ and $n_2$ are phase winding numbers in the upper and lower arms of the ring, respectively. (c) and (d) Calculated (c) SC transition temperature and (d) critical bias current obtained for the spin-singlet $s$-wave ring of radius $R/\xi_0=50$ in the settings (a) and (b), respectively, where colored curves are obtained for given $n$'s or sets of $n_1$ and $n_2$, and their largest values are traced by a dotted black curve which exhibits the LP oscillation. (d) is obtained at $T/T_{c0}=0.6$. \label{fig:fig1}}
\end{center}
\end{figure}

An experimental setup for the usual LP oscillation in $T_c$ is illustrated in Fig. \ref{fig:fig1} (a). Assuming that the ring width $w$ and thickness $d$ are sufficiently small, the SC gap function for the spin-singlet $s$-wave pairing is expressed as $\Delta_s =|\Delta|e^{-i n \varphi}$ with integer winding number $n$ and azimuthal angle $\varphi$. In the presence of an axial magnetic field ${\bf H}$, the SC current circulating around the ring is given by ${\bf j}\propto \hat{\varphi}|\Delta|^2(n-\frac{\Phi}{\Phi_0})$ with the magnetic flux passing through the ring $\Phi$ \cite{LP_ParksLittle_pr_64, Thinkham}. Here, the first and second terms in ${\bf j}$ originate from the phase gradient and the London screening, respectively. With increasing field, or equivalently, increasing $\Phi$, the integer $n$ switches to a larger value to reduce ${\bf j}$. The occurrence of such a higher-$n$ state and the associated discontinuous change in $n$ are reflected in $T_c$ as the LP oscillation [see Fig. \ref{fig:fig1} (c) and the text below] \cite{LP_LittleParks_prl_62,LP_ParksLittle_pr_64,Thinkham}. 

On the other hand, as illustrated in Fig. \ref{fig:fig1} (b), when a bias current is applied to the ring from left to right, upper and lower halves of the ring are not equivalent any more, and the properties of this current-carrying SC state are not so trivial \cite{LP-jc_Gurtovoi_jetp_07,LP-jc_Gurtovoi_pla_20,LP-jc_Michotte_prb_10}. Furthermore, in a recent LP experiment done on polycrystalline $\beta$-Bi$_2$Pd in this setting, a half-quantum-shifted LP oscillation has been observed and the realization of a spin-triplet SC state has been discussed \cite{Bi2Pd_Li_science_19}. In view of such a situation, we theoretically investigate the LP oscillation in the presence of the bias current for both the spin-singlet and spin-triplet SC rings without any domains and Josephson junctions.

In the LP experiment with the bias current, one usually measures the resistivity which is associated with the critical bias current $j_c$. 
In this work, we calculate $j_c$ at a low temperature below $T_c$ where the SC gap is well developed. Based on the Ginzburg-Landau (GL) theory, we show that in the spin-triplet case, not only the phase but also the ${\bf d}$-vector rotates along the circumference of the ring to yield a larger $j_c$. To our knowledge, there is no well-established theory of $j_c$ even for the spin-singlet SC ring ($j_c$ derived in Refs. \cite{LP-jc_Gurtovoi_pla_20,LP-jc_Michotte_prb_10} will be discussed later on), so that we shall start from the LP oscillation for the spin-singlet $s$-wave ring.     

{\it Spin}-{\it singlet} {\it s}-{\it wave} {\it case.}--
For the bulk singlet $s$-wave superconductor, the GL free energy ${\cal F}_{\rm GL}$ and the SC current ${\bf j}$ are derived from the weak-coupling BCS Hamiltonian as ${\cal F}_{\rm GL} = \int d{\bf r} \big[ \alpha |\Delta_s|^2 + \beta |\Delta_s|^4 + K |{\boldsymbol\Pi} \Delta_s |^2 \big]$ and ${\bf j} = - \frac{\delta {\cal F}_{\rm GL}}{\delta {\bf A}}=-2|e|K (\Delta_s^\ast {\boldsymbol\Pi} \Delta_s + \Delta_s {\boldsymbol\Pi}^\dagger \Delta_s^\ast )$, where the coefficients are given by $\alpha=N(0)\ln\frac{T}{T_{c0}}$, $\beta=\frac{7\zeta(3) N(0)}{16 \pi^2 T^2}$, and $K=N(0) (\frac{T_{c0}}{T})^2 \, \xi_0^2$ with density of states at the Fermi level $N(0)$, the SC transition temperature at zero field $T_{c0}$, and the SC coherence length $\xi_0=\sqrt{\frac{7\zeta(3)}{48 \pi^2}} \, \frac{v_{\rm F}}{T_{c0}}$ \cite{Parks}. For the SC ring with sufficiently small width $w$ and thickness $d$, the gap function basically depends only on $\varphi$, namely, $\Delta_s =|\Delta|e^{-i n \varphi}$, and the gauge field can be expressed as ${\bf A}=\frac{1}{2}{\bf H} \times {\bf r} = \frac{H \, R}{2} {\hat \varphi}$, so that ${\boldsymbol\Pi}=-i {\boldsymbol\nabla}+2|e|{\bf A}$ is reduced to ${\boldsymbol\Pi}={\hat \varphi}\frac{1}{R}\big( -i\partial_{\varphi} + \frac{\Phi}{\Phi_0}\big)$ with $\Phi=\pi R^2 H$.

By substituting $\Delta_s =|\Delta|e^{-i n \varphi}$ into ${\cal F}_{\rm GL}$, one obtains the free energy density $f_{\rm GL}={\cal F}_{\rm GL}/(2\pi R w d)$ as $f_{\rm GL}= |\Delta|^2\big[\alpha+\frac{K}{R^2}(n-\frac{\Phi}{\Phi_0})^2\big] + \beta |\Delta|^4$. Since the SC transition temperature is determined by the GL quadratic term via the condition $\alpha+\frac{K}{R^2}(n-\frac{\Phi}{\Phi_0})^2=0$, it turns out that with increasing $\Phi$, the phase winding number $n$ switches to a larger value to lower the gradient energy $\frac{K}{R^2}(n-\frac{\Phi}{\Phi_0})^2$.  
The calculation result for $R/\xi_0=50$ is shown in Fig. \ref{fig:fig1} (c). One can see that the highest transition temperature indicated by a dotted curve in Fig. \ref{fig:fig1} (c) exhibits a quantum oscillation with its period being characterized by $\Phi_0$, i.e., the LP oscillation. We note in passing that roles of larger $w$ and $d$ have been discussed elsewhere \cite{LP_Groff_pr_68, LP_ABSV_prl_13}. Also, in quantum-wire systems where electrons are confined in nanostructures, a ring shape seems to be important \cite{Qwire_Cuoco_17}, but such a shape effect should be irrelevant in the present LP system with $w, \, d \gtrsim \xi_0$.

In the case with the bias current shown in Fig. \ref{fig:fig1} (b) , the phases in the upper and lower arms of the ring are not necessarily the same, so that we assume 
\begin{equation}\label{eq:OP_s}
\Delta_s = |\Delta| \left\{\begin{array}{l}
e^{-in_{1} \varphi} \, \, (0\leq \varphi \leq \pi) \\
e^{-in_{2} \varphi} \, (\pi \leq \varphi \leq 2\pi) \\
\end{array} \right.  
\end{equation}
with integers $n_{1}$ and $n_{2}$. Then, one notices that $n_{1}-n_{2}$ must be an even integer such that $\Delta_s$ be nonzero everywhere; if $n_{1}-n_{2}$ is an odd integer, $\displaystyle{\lim_{\varphi \rightarrow \pi+0}}\Delta_s$ and $\displaystyle{\lim_{\varphi \rightarrow \pi-0}}\Delta_s$ take opposite signs, so that $|\Delta|=0$ at $\varphi=\pi$. The free energy density $f_{1}$ ($f_{2}$) and the SC current ${\bf j}_{1}$ (${\bf j}_{2}$) in the upper (lower) arm are expressed as 
\begin{eqnarray}\label{eq:halfring_s}
f_{\,l} &=& \alpha|\Delta|^2 + \beta |\Delta|^4 + |\Delta|^2 \frac{K}{R^2}\Big(n_l-\frac{\Phi}{\Phi_0}\Big)^2, \nonumber\\
{\bf j}_{\, l} &=& (-1)^{l} 4|e|K |\Delta|^2 \frac{1}{R}\Big(n_{l}-\frac{\Phi}{\Phi_0} \Big) \hat{\varphi},
\end{eqnarray} 
where the additional minus sign enters in ${\bf j}_1$ as it is defined in the clockwise direction in Fig. \ref{fig:fig1} (b).
The total free energy density $f_{\rm GL}$ and the total bias current ${\bf j}_{\rm tot}$ are given by $f_{\rm GL}=(f_1+f_2)/2$ and ${\bf j}_{\rm tot}={\bf j}_1+{\bf j}_2$. 

Now, we calculate the critical bias current $j_c$ inside the SC phase at $T<T_{c0}$, i.e., $\alpha<0$, in the same procedure as that to derive the critical current in one dimension $j_c^{\rm (1D)}$ \cite{Thinkham}. Since the amplitude of the SC gap function $|\Delta|$ is determined from the condition $\frac{\delta f_{\rm GL}}{\delta |\Delta|}=0$ as
\begin{eqnarray}
|\Delta|^2 &=& \frac{-1}{2\beta}\Big[ \alpha + \frac{K}{2R^2}\Big\{\Big(n_1-\frac{\Phi}{\Phi_0}\Big)^2 + (n_2-\frac{\Phi}{\Phi_0}\Big)^2\Big\}\Big] \nonumber\\
&=& \frac{-1}{2\beta}\Big[ \alpha + \frac{K}{R^2}\Big\{ \Big(n_2 - \frac{n_{\rm tot}}{2} - \frac{\Phi}{\Phi_0} \Big)^2+ \frac{n_{\rm tot}^2}{4} \Big\} \Big]
\end{eqnarray}
with $n_{\rm tot}=n_2-n_1$, ${\bf j}_{\rm tot}$ is expressed as 
\begin{equation}\label{eq:jc_s-wave}
|{\bf j}_{\rm tot}|= \frac{2|e|K}{\beta R} \Big[ |\alpha| - \frac{K}{R^2}\Big\{ \Big(n_2 - \frac{n_{\rm tot}}{2} - \frac{\Phi}{\Phi_0} \Big)^2+ \frac{n_{\rm tot}^2}{4} \Big\} \Big] \big| n_{\rm tot} \big|.
\end{equation}
Then, $j_c$ is obtained as the maximum value of $|{\bf j}_{\rm tot}|$ as a function of integers $n_1$ and $n_2$ under the constraint of $n_{\rm tot}=n_2-n_1$= even integer, or equivalently, arbitrary integers $n_2$ and $n_{\rm tot}/2$. Figure \ref{fig:fig1} (d) shows a typical example of the calculated $j_c$ as a function of $\Phi$, where $j_c$ is normalized by $j_0=N(0)T_{c0} \, |e|v_{\rm F}$. Among $j_c$'s for various combinations of $n_1$ and $n_2$, the largest one gives the physical critical current which is indicated by a dotted curve in Fig. \ref{fig:fig1} (d). One can see the LP oscillation in $j_c$ similar to that in $T_c$.  

In Refs. \cite{LP-jc_Gurtovoi_pla_20, LP-jc_Michotte_prb_10}, $j_c$ is derived in a different way, where it is assumed that the bias current is split equally into two and one arm with a larger net current consisting of the bias and additional circulating currents [suppose that it is, for example, the upper arm in Fig. \ref{fig:fig1} (b)] determines $j_c$ by the condition $|{\bf j}_1|=j_c^{\rm (1D)}$, leading to $j_c$ showing a triangle-wave function of $\Phi$. In the opposite arm, however, $|{\bf j}_2|<j_c^{\rm (1D)}$ should be satisfied because the circulating current cancels the bias current, so that the half of the system remains SC. In contrast, $j_c$ obtained from Eq. (\ref{eq:jc_s-wave}) [see Fig. \ref{fig:fig1} (d)] determines the SC instability over the whole system. 
In Al rings, the LP oscillation in $j_c$, which looks rather similar to Fig. \ref{fig:fig1} (d), has been observed \cite{LP-jc_Gurtovoi_jetp_07,LP-jc_Gurtovoi_pla_20}, validating the present approach, although its detailed structure depends on ring asymmetries \cite{LP-jc_Gurtovoi_pla_20}. In Nb rings, the $j_c$ oscillation has been observed as the LP oscillation in the resistivity \cite{LP-jc_Tokuda_jjap_22}, and for a specific ring shape, vortex penetrations seem to affect $j_c$ \cite{LP-jc_Michotte_prb_10}. 

{\it Spin}-{\it triplet} {\it p}-{\it wave} {\it case.}--
Next, we will discuss the spin-triplet case where the gap function is expressed as
\begin{equation}
\hat{\Delta}_t=\left( \begin{array}{cc}
\Delta_{\uparrow\uparrow} & \Delta_{\uparrow\downarrow} \\
\Delta_{\downarrow\uparrow} & \Delta_{\downarrow\downarrow}
\end{array} \right) = \left( \begin{array}{cc}
-d_x+id_y & d_z \\
d_z & d_x+id_y
\end{array} \right)  
\end{equation}
with the ${\bf d}$-vector ${\bf d}=(d_x,d_y,d_z)$. In a magnetic field, the $\uparrow\downarrow$ Cooper pair becomes unfavorable, so that the ${\bf d}$-vector tends to orient perpendicularly to the field, i.e., $d_z=0$. For the ring geometry without the bias current, the ${\bf d}$-vector can be expressed with the use of two winding numbers $m$ and $n$ as 
\begin{equation}\label{eq:HQV_general}
{\bf d} = \Big[ \frac{|\Delta_{\uparrow\uparrow}|}{\sqrt{2}} {\hat d}^+ \, e^{-i m \varphi} + \frac{|\Delta_{\downarrow\downarrow}|}{\sqrt{2}} {\hat d}^- \, e^{im \varphi}\Big] \, e^{-i n \varphi},
\end{equation}
where ${\hat d}^\pm =\frac{1}{\sqrt{2}}({\hat x} \pm i \, {\hat y})$. To grasp the physical meanings of $m$ and $n$, it is convenient to consider the simplified case of $|\Delta|=|\Delta_{\uparrow\uparrow}|=|\Delta_{\downarrow\downarrow}|$ where Eq. (\ref{eq:HQV_general}) is reduced to ${\bf d} = |\Delta| \, e^{-i n \varphi} \, \big[ {\hat x}\cos(m \varphi) + {\hat y}\sin(m \varphi) \big]$. It is clear that $n$ is the phase winding number, whereas $m$ describes the rotation of the ${\bf d}$-vector along the circumference of the ring.  
In the case of $m=0$, the ${\bf d}$-vector is spatially uniform and $n$ takes an integer value as in the spin-singlet case. For $m \neq 0$, on the other hand, $n$ and $m$ can take half-integer values as well as integer values. Since the gap function should be single-valued, $n$ must be a half integer for a half-integer value of $m$ to avoid the sign change at the two equivalent points of $\varphi=0$ and $2\pi$. This half-integer state accompanied with the ${\bf d}$-vector rotation corresponds to the so-called half-quantum vortex which has often been discussed in the context of superfluid $^3$He \cite{HQV_ABM_Salomaa_prl_85, HQV_polar_Autti_prl_16, HQV_review_Sals_16, HQV_polar_Nagamura_prb_18}. 

In the spin-triplet superconductor with the isotropic $p$-wave pairing interaction, the ${\bf d}$-vector is expressed as $d_\mu = \sum_i A_{\mu,i}\hat{p}_i$ with the order parameter $A_{\mu,i}$, and the GL free energy is given by ${\cal F}_{\rm GL}=\int d{\bf r}\big[ f^{(2)}+\delta f^{(2)}+f^{(4)}\big]$ with
\begin{eqnarray}\label{eq:GL_t}
f^{(2)} &=& \sum_{\mu,i,j} A^\ast_{\mu,i}\Big[ \frac{\alpha}{3}\delta_{i,j}A_{\mu,j} + K'\big( \Pi_j\Pi_j A_{\mu,i} + 2 \Pi_i \Pi_j A_{\mu,j} \big)\Big], \nonumber\\
\delta f^{(2)} &=& - \frac{\delta \alpha}{3} i \, \sum_{i} (A_{x,i}A^\ast_{y,i}-A_{y,i}A^\ast_{x,i}), \nonumber\\
f^{(4)} &=& \sum_{\mu,\nu,i,j} \big(\beta_1 |A_{\mu,i}A_{\mu,i}|^2+\beta_2(A^\ast_{\mu,i}A_{\mu,i})^2 \nonumber\\
&& + \beta_3 A^\ast_{\mu,i}A^\ast_{\nu,i}A_{\mu,j}A_{\nu,j} + \beta_4 A^\ast_{\mu,i}A_{\nu,i}A^\ast_{\nu,j}A_{\mu,j} \nonumber\\
&& + \beta_5 A^\ast_{\mu,i}A_{\nu,i}A_{\nu,j}A^\ast_{\mu,j} \big).
\end{eqnarray}
This functional form is obtained from ${\cal F}_{\rm GL}$ for superfluid $^3$He by replacing $-i {\boldsymbol\nabla}$ with ${\boldsymbol\Pi}$, so that the coefficients are given by $-2\beta_1 = \beta_2=\beta_3=\beta_4=-\beta_5 = 2\beta_0$ with $\beta_0 = \frac{7\zeta(3)N(0)}{240 \pi^2T^2}$ and $K'=\frac{1}{5}N(0)(\frac{T_{c0}}{T})^2 \, \xi_0^2$ in the weak-coupling limit \cite{VW}. The $\delta f^{(2)}$ term originates from the difference in density of states between the up-spin and down-spin Fermi surfaces and thereby, is active only in a magnetic field \cite{VW,HQV_Vakaryuk_prl_09}. Within a linear approximation, we could express $\delta \alpha$ as $\delta \alpha = N(0) \, \eta \frac{\Phi}{\Phi_0}$. The SC current ${\bf j}=-\frac{\delta {\cal F}_{\rm GL}}{\delta {\bf A}}$ is given by $j_\alpha = -4|e| K' \sum_{\mu, i,j} A^\ast_{\mu,i}(\delta_{ij}\Pi_\alpha +\delta_{\alpha j}\Pi_i+\delta_{\alpha i}\Pi_j) A_{\mu,j}$.

\begin{figure}[t]
\begin{center}
\includegraphics[scale=0.62]{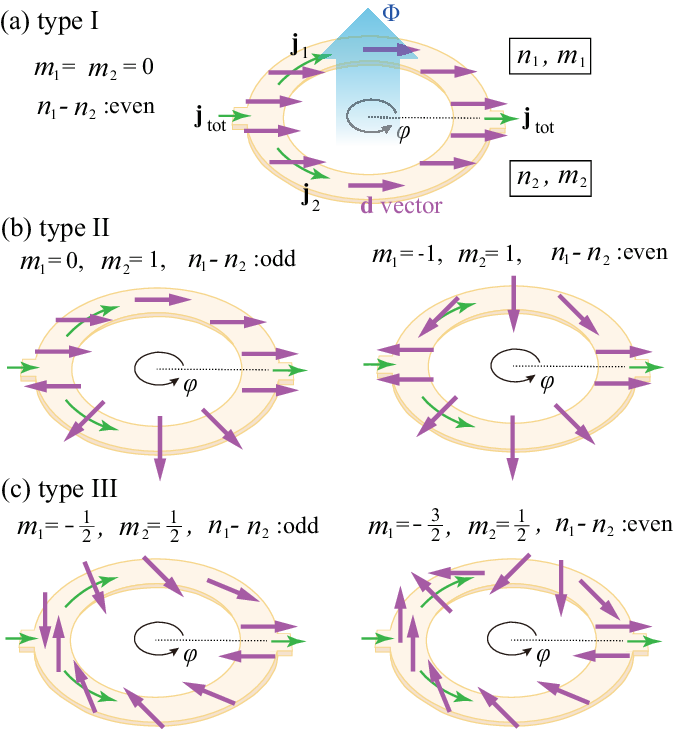}
\caption{Schematically drawn ${\bf d}$-vector textures in the spin-triplet SC ring in the presence of both the magnetic flux $\Phi$ and the bias current ${\bf j}_{\rm tot}$, where a magenta arrow represents the ${\bf d}$-vector and $|\Delta_{\uparrow\uparrow}|=|\Delta_{\downarrow\downarrow}|$ is assumed for ease of understanding. (a) Type-I state with $m_1=m_2=0$, (b) type-II states with $m_1=0$ and $m_2=1$ (left) and $m_1=-1$ and $m_2=1$ (right), and (c) type-III states with $m_1=-\frac{1}{2}$ and $m_2=\frac{1}{2}$ (left) and $m_1=-\frac{3}{2}$ and $m_2=\frac{1}{2}$ (right), which are categorized in Table \ref{table:constraint}.  \label{fig:fig2}}
\end{center}
\end{figure}
\begin{table}[b]
\caption{Constraint on the phase winding numbers $n_1$ and $n_2$ and the ${\bf d}$-vector winding numbers $m_1$ and $m_2$. \label{table:constraint}}

\begin{tabular}{|c|c|c|c|c|}
\hline
type & $n_1$, $n_2$ & $m_1$, $m_2$ & $n_1-n_2$ & $m_1-m_2$ \\
\hline
I & integer & $m_1=m_2=0$  & even & 0    \\ 
\hline
\multirow{2}{*}{II} & \multirow{2}{*}{integer} & \multirow{2}{*}{integer$\neq 0$} & even & even   \\ 
& & & odd & odd        \\
\hline
\multirow{2}{*}{III} & \multirow{2}{*}{half integer} & \multirow{2}{*}{half integer} & even & even    \\ 
& & & odd & odd       \\
\hline
\end{tabular}
\end{table} 
Now, we consider a possible form of $A_{\mu,i}$ for the ring geometry. For simplicity, we assume that the spin and orbital degrees of freedom are decoupled such that $A_{\mu,i}=a_\mu \, w_{{\bf p},i}$, and take an axially-symmetric orbital state of the form $w_{{\bf p},i}=\frac{1}{\sqrt{2}}(\hat{p}_x+i \, \hat{p}_y)_i$. 
We note that even if the polar state of the axially-symmetric form $w_{{\bf p},i}=\hat{p}_{z,i}$ is assumed, the following discussion is essentially the same except the overall prefactors, so that the chiral nature of the orbital pairing state is not important. 

Concerning the spin degrees of freedom $a_\mu$, Eq. (\ref{eq:HQV_general}) can be straightforwardly extended to the case with the bias current, and $a_\mu $ can be expressed as
\begin{equation}\label{eq:OP_t}
a_\mu = \left\{\begin{array}{l}
\big[\frac{|\Delta_{\uparrow\uparrow}|}{\sqrt{2}} {\hat d}^+_\mu \, e^{-i m_1 \varphi}+ \frac{|\Delta_{\downarrow\downarrow}|}{\sqrt{2}} {\hat d}^-_\mu \, e^{i m_1 \varphi}\big] e^{-i n_1 \varphi}\, (0\leq \varphi \leq \pi) \nonumber\\
\big[\frac{|\Delta_{\uparrow\uparrow}|}{\sqrt{2}} {\hat d}^+_\mu \, e^{-i m_2 \varphi} + \frac{|\Delta_{\downarrow\downarrow}|}{\sqrt{2}}{\hat d}^-_\mu \, e^{i m_2 \varphi}\big]  e^{-i n_2 \varphi} \, (\pi \leq \varphi \leq 2\pi) 
\end{array} \right. .
\end{equation}
As in the spin-singlet case, there exists a constraint on $n_1, \, n_2, \, m_1$, and $m_2$ such that $|\Delta_{\uparrow\uparrow}|$ and $|\Delta_{\downarrow\downarrow}|$ be nonzero even at the intersections between the upper and lower arms, namely, at $\varphi=0$ and $\pi$; $n_1-n_2$ must be an even (odd) integer when $m_1-m_2$ is an even (odd) integer, which is summarized in Table \ref{table:constraint}. In Table \ref{table:constraint}, type I represents the $m_1=m_2=0$ state where the ${\bf d}$-vector is spatially uniform [see Fig. \ref{fig:fig2} (a)], whereas types II and III represent the states with $m_1 \neq 0$ and/or $m_2 \neq 0$ where the ${\bf d}$-vector forms a texture along the circumference [see Figs. \ref{fig:fig2} (b) and (c)]. In the type-II (type-III) state, $n_1$, $n_2$, $m_1$, and $m_2$ are integers (half integers), and thus, as illustrated in Fig. \ref{fig:fig2} (b) [(c)], the ${\bf d}$-vectors at $\varphi=0$ and $2\pi$ are parallel (antiparallel) to each other. 

By substituting $A_{\mu, i} = a_\mu \, w_{{\bf p},i}$ into Eq. (\ref{eq:GL_t}), we obtain the total free energy density $f_{\rm GL}=(f_1+f_2)/2$ as
\begin{eqnarray}\label{eq:energy_tot_t}
&& f_{\rm GL} = \frac{1}{2}\sum_{\sigma=\uparrow, \downarrow}\Big\{\frac{\alpha-\epsilon_{\sigma} \, \delta \alpha}{3}|\Delta_{\sigma\sigma}|^2 + 2\beta_0 |\Delta_{\sigma\sigma}|^4 \\
&& + \frac{K'}{R^2}|\Delta_{\sigma\sigma}|^2\Big[\big(n_1+\epsilon_\sigma m_1-\frac{\Phi}{\Phi_0}\big)^2+\big(n_2+\epsilon_\sigma m_2-\frac{\Phi}{\Phi_0}\big)^2 \Big]\Big\} \nonumber ,
\end{eqnarray}
where $\epsilon_{\uparrow (\downarrow)}=1 \, (-1)$. In the same manner, the total current ${\bf j}_{\rm tot}={\bf j}_1+{\bf j}_2$ is calculated as
\begin{eqnarray}\label{eq:current_tot_t}
{\bf j}_{\rm tot} &=&-{\hat \varphi} \, \frac{4|e|K'}{R}\Big[(|\Delta_{\uparrow\uparrow}|^2+|\Delta_{\downarrow\downarrow}|^2)(n_1-n_2) \nonumber\\
&+& (|\Delta_{\uparrow\uparrow}|^2-|\Delta_{\downarrow\downarrow}|^2)(m_1-m_2) \Big]
\end{eqnarray}
with
\begin{eqnarray}\label{eq:GLeq_t}
|\Delta_{\sigma\sigma}|^2 &=& \frac{-1}{4\beta_0}\Big\{ \frac{\alpha-\epsilon_{\sigma} \, \delta \alpha}{3} + \frac{K'}{R^2}\Big[\big(n_1+\epsilon_\sigma m_1-\frac{\Phi}{\Phi_0}\big)^2 \nonumber\\
&+& \big(n_2+\epsilon_\sigma m_2-\frac{\Phi}{\Phi_0}\big)^2 \Big] \Big\},
\end{eqnarray}
where Eq. (\ref{eq:GLeq_t}) is obtained from the conditions $\frac{\delta f_{\rm GL}}{\delta |\Delta_{\uparrow\uparrow}|}=0$ and $\frac{\delta f_{\rm GL}}{\delta |\Delta_{\downarrow\downarrow}|}=0$.
The maximum value of $|{\bf j}_{\rm tot}|$ as a function of $n_1$, $n_2$, $m_1$, and $m_2$ under the constraints shown in Table \ref{table:constraint} corresponds to $j_c$. 

\begin{figure}[t]
\begin{center}
\includegraphics[scale=0.58]{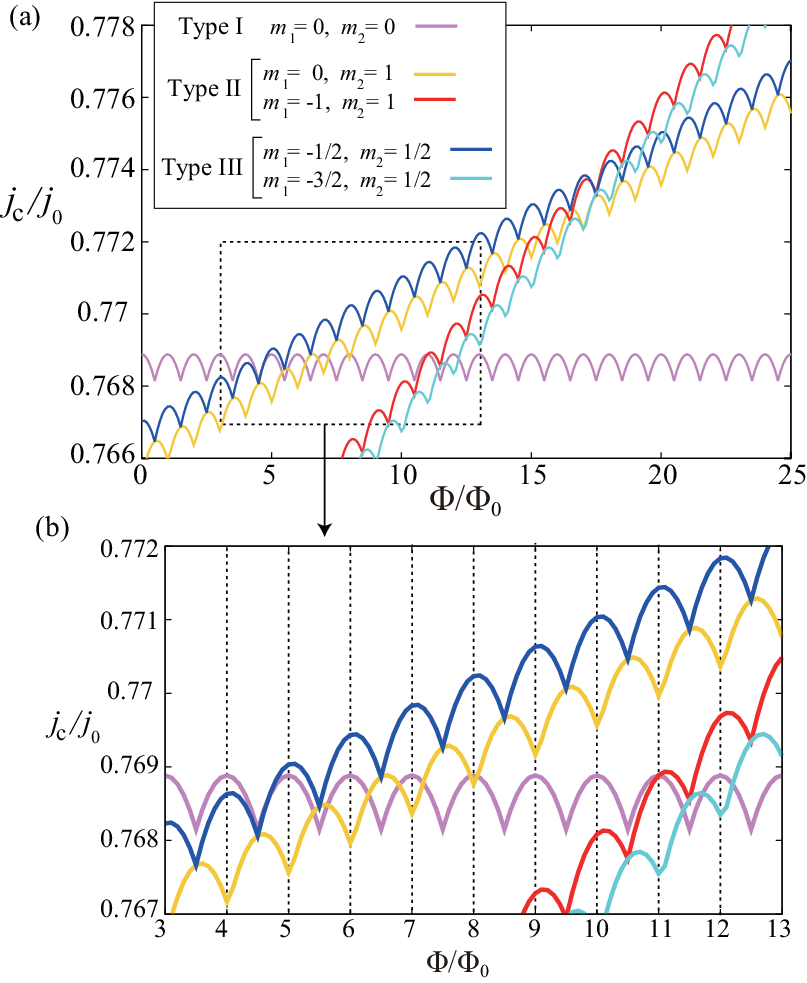}
\caption{LP oscillation in $j_c$ obtained at $T/T_{c0}=0.6$ for the spin-triplet $p$-wave ring with $R/\xi_0=50$ and $\eta=0.004$. (a) $j_c$'s for the five types of ${\bf d}$-vector textures shown in Fig. \ref{fig:fig2}; violet (type I with $m_1=m_2=0$), yellow (type II with $m_1=0,m_2=1$), red (type II with $m_1=-1,m_2=1$), blue (type III with $m_1=-\frac{1}{2},m_2=\frac{1}{2}$), and cyan (type III with $m_1=-\frac{3}{2},m_2=\frac{1}{2}$). A zoomed view of the region enclosed by a dotted box is shown in (b). In (b), dashed lines indicate peak positions of the LP oscillation for the type-I state realized at low fields. \label{fig:fig3}}
\end{center}
\end{figure}

Figure \ref{fig:fig3} shows the calculation result for $R/\xi_0=50$, $T/T_{c0}=0.6$, and $\eta=0.004$. 
As one can see from Fig. \ref{fig:fig3} (a), $j_c$ for the type-I state (violet curve) exhibits the conventional LP oscillation, whereas $j_c$'s for the type-II (red and yellow curves) and type-III (blue and cyan curves) states are increasing functions of $\Phi$ accompanied with the LP oscillation. The largest critical current is given by the type-I state at low fields, whereas at higher fields, it is given by the type-II or type-III state with $|m_1|=|m_2|\neq 0$ [blue or red curve in Fig. \ref{fig:fig2} (a)], suggesting that the transition between the uniform and textured ${\bf d}$-vector configurations occurs.  
In Fig. \ref{fig:fig3}, $j_c$'s for $|m_1|\neq |m_2|$ (yellow and cyan curves) are always smaller than $j_c$'s for $|m_1|=|m_2|$ (red and blue curves) except degenerate points, so that the $|m_1|\neq |m_2|$ states are not realized at least within the realm of this theory. 

As one can see from the zoomed view shown in Fig. \ref{fig:fig3} (b), the phase of the LP oscillation in the physical critical current, i.e., the largest $j_c$, remains unchanged without showing a half-quantum shift, even after the transition into the type-III state with the ${\bf d}$-vector texture (compare the violet and blue curves). In the case of $m_1=0$ and $m_2=1$ (see a yellow curve), on the other hand, the oscillation pattern is half-quantum shifted compared with the low-field conventional LP oscillation, although such a half-quantum shifted state with $|m_1|\neq |m_2|$ is unstable.

Now, we discuss why the ${\bf d}$-vector texture yields the larger $j_c$ in the presence of the magnetic field. 
In Eq. (\ref{eq:current_tot_t}), one notices that ${\bf j}_{\rm tot}$ consists of two parts: one is a phase-gradient current proportional to $n_1-n_2$, and the other is a ${\bf d}$-vector-texture-induced one proportional to $m_1-m_2$. In the type-I state with $m_1=m_2=0$, ${\bf j}_{\rm tot}$ is governed only by the former which is coupled to $|\Delta_{\uparrow\uparrow}|^2+|\Delta_{\downarrow\downarrow}|^2$. Since the effect of the Zeeman splitting, $\pm\delta \alpha$, is completely cancelled out in $|\Delta_{\uparrow\uparrow}|^2+|\Delta_{\downarrow\downarrow}|^2$ [see Eq. (\ref{eq:GLeq_t})], $j_c$ for the type-I state is field-independent except the LP oscillation. In contrast, in the type-II and type-III states with $m_1-m_2 \neq 0$, the latter ${\bf d}$-vector-texture term is active being coupled to $|\Delta_{\uparrow\uparrow}|^2-|\Delta_{\downarrow\downarrow}|^2$ which is roughly proportional to $\delta \alpha = \eta \frac{\Phi}{\Phi_0}$, so that $j_c$ increases with increasing field. Since the ${\bf d}$-vector texture itself costs the gradient energy, the type-II and type-III states are unstable at zero field, but with increasing field, their $j_c$' are elevated by the ${\bf d}$-vector-texture term proportional to both $m_1-m_2$ and $\eta \frac{\Phi}{\Phi_0}$, eventually overwhelming the type-I value. Thus, in a system with a larger Fermi surface Zeeman splitting $\eta$, the transition between the type-I and type-III states occurs at a lower field.     

The above discussion on $j_c$ can be understood from the viewpoint of the free energy. With the use of $n_{\rm tot}=n_2-n_1$, the gradient energy in Eq. (\ref{eq:energy_tot_t}) can be rewritten as 
\begin{eqnarray}\label{eq:grad}
&&\frac{K'}{R^2} \big(|\Delta_{\uparrow\uparrow}|^2+|\Delta_{\downarrow\downarrow}|^2 \big) \Big[\Big(n_2-\frac{n_{\rm tot}}{2}-\frac{\Phi}{\Phi_0}+\lambda\frac{m_1+m_2}{2}\Big)^2 \nonumber\\
&& +\frac{n_{\rm tot}^2}{4}+\frac{m_1^2}{2}+\frac{m_2^2}{2} -  \lambda \, n_{\rm tot} \frac{m_1-m_2}{2} -\Big(\lambda \frac{m_1+m_2}{2}\Big)^2\Big] \nonumber
\end{eqnarray}
with $\lambda = \frac{|\Delta_{\uparrow\uparrow}|^2-|\Delta_{\downarrow\downarrow}|^2}{|\Delta_{\uparrow\uparrow}|^2+|\Delta_{\downarrow\downarrow}|^2}$.
Due to the $- \lambda  \,n_{\rm tot} \, (m_1-m_2)$ term, which is active in the presence of both the bias current ${\bf j}_{\rm tot}$ roughly proportional to $n_{\rm tot} \neq 0$, and the Zeeman splitting, i.e., $\lambda\propto \eta\frac{\Phi}{\Phi_0}\neq 0$, the ${\bf d}$-vector texture with $m_1-m_2 \neq 0$ can acquire the larger energy gain than the energy cost of $m_1^2+m_2^2$, leading to the occurrence of the type-II and type-III states. We note that such a linear coupling between the magnetic field and the SC current occurs in the different context of noncentrosymmetric superconductors \cite{NCS_book} where the coupling yields a phase-modulated helical SC state and associated magnetoelectric phenomena \cite{SC_jM,SC_Bj,Dimitrova,Samokhin,Kaur,Fujimoto,Helical_AS_12,Helical_ASS_14}. 

$Discussion.$--
In this letter, we have demonstrated that in the spin-triplet SC ring, the critical bias current exhibits the LP oscillation, being accompanied by the field-induced transition into the textured state of the ${\bf d}$-vector. Here, we comment on additional effects which are not incorporated in the present GL analysis. First, we have assumed the discontinuous change in winding numbers at the intersections between the upper and lower arms whose gradient energy may suppress the stability of the textured state. 
Such an additional energy cost, however, could be estimated by using the functional form near $\varphi=0$, $a_\mu = |\Delta_{\sigma\sigma}|e^{-i \, m(\varphi) \, \varphi}$ with $m(\varphi)=\frac{1}{2}\big[ m_1+m_2 +(m_1-m_2)\tanh(\varphi/L_0)\big]$ and $L_0=\xi_0/R$, as $\int_{-L_0}^{L_0}K'|\nabla a_{\mu}|^2 \, d\varphi \sim \frac{K'}{R^2}|\Delta_{\sigma\sigma}|^2 \, \frac{\xi_0}{R}$, so that for a relatively large ring radius, it should be irrelevant compared with the energy gain $-\frac{K'}{R^2}|\Delta_{\sigma\sigma}|^2 \, \lambda \,n_{\rm tot} \,  (m_1-m_2)$ to stabilize the ${\bf d}$-vector texture. 
Second, although our analysis is based on the weak-coupling BCS theory, the Fermi liquid correction may assists the ${\bf d}$-vector texture as in the case of superfluid $^3$He \cite{HQV_ABM_Salomaa_prl_85, HQV_polar_Nagamura_prb_18, HQV_Vakaryuk_prl_09}. Third, we have assumed that the spin space is isotropic and the ${\bf d}$-vector can rotate freely, but in real materials, the ${\bf d}$-vector may be subject to pinning due to, for example, magnetic impurities. 

In the presence of the ${\bf d}$-vector pinning, when the field is gradually increased in the initial uniform state shown in Fig. \ref{fig:fig2} (a), the type-II state possessing the half-uniform and half-textured ${\bf d}$-vector configuration shown in the left panel of Fig. \ref{fig:fig2} (b) might possibly be realized with the ${\bf d}$-vector in the upper arm being pinned to be almost uniform, instead of the theoretically expected type-III state with the fully textured ${\bf d}$-vector configuration shown in the left panel of Fig. \ref{fig:fig2} (c). If it happens, a switching from the usual LP oscillation to the half-quantum-shifted one [see the yellow curve in Fig. \ref{fig:fig3} (b)] should be observed, being distinguished from the domain-induced unconventional LP pattern which is half-quantum shifted from the beginning \cite{Bi2Pd_Li_science_19, Bi2Pd_Xu_prl_20}. Although such a field-induced half-quantum shift has not been reported so far even in the LP experiments conducted for the spin-triplet candidate Sr$_2$RuO$_4$ \cite{Sr2RuO4-LP_Cai_prb_13,Sr2RuO4-LP_Yasui_prb_17,Sr2RuO4-LP_Yasui_npj_20,Sr2RuO4-LP_Cai_arXiv_20} (its pairing symmetry is still controversial \cite{Sr2RuO4_NMR_Pustogow_nature_19}), we believe that our result presented here will promote the exploration and further understanding of not only spin-triplet but also various classes of superconductors.

\begin{acknowledgments}
The author thanks M. Tokuda and Y. Niimi for stimulating discussions and T. Mizushima, R. Ikeda, and M. Sigrist for valuable comments. This work is partially supported by JSPS KAKENHI Grants No. JP21K03469.
\end{acknowledgments}

\end{document}